\documentclass[aps,preprintnumbers,nofootinbib,floatfix]{revtex4}
\usepackage{amssymb,bm,graphicx}

\begin{document}

\title{The relation between TMDs and PDFs in the covariant parton model approach}
\author{A.V.~Efremov$^{1}$, P. Schweitzer$^{2}$, O.~V.~Teryaev$^{1}$, P.~Zavada$^{3}$}
\affiliation{{\ }$^{1}$ {Bogoliubov Laboratory of Theoretical Physics, JINR, 141980
Dubna, Russia}\\
$^{2}$ {Department of Physics, University of Connecticut, Storrs, CT 06269,
U.S.A.}\\
$^{3}$ {Institute of Physics AS CR, Na Slovance 2, CZ-182 21 Prague 8, Czech
Rep.}}

\begin{abstract}
We derive relations between transverse momentum dependent distribution functions
(TMDs) and the usual parton distribution functions (PDFs) in the 3D covariant 
parton model, which follow from Lorentz invariance and the assumption of a 
rotationally symmetric distribution of parton momenta in the nucleon rest frame. 
Using the known PDFs $f_1^q(x)$ and $g_1^q(x)$ as input we predict the $x$- 
and ${\bf p}_T$-dependence of all twist-2 T-even TMDs.
\end{abstract}

\maketitle

\section{Introduction}
\label{sec1}

TMDs \cite{tmds,Mulders:1995dh} open a new way to a more complete
understanding of the quark-gluon structure of the nucleon. Indeed, 
some experimental observations can hardly be explained without a more 
accurate and realistic 3D picture of the nucleon, which naturally includes 
transverse motion. The azimuthal asymmetry in the distribution of hadrons 
produced in deep-inelastic lepton-nucleon scattering (DIS), known as the Cahn 
effect \cite{Cahn:1978se}, is a classical example. 
The intrinsic (transversal) parton motion is also
crucial for the explanation of some spin effects
\cite{Airapetian:1999tv,Avakian:2003pk,Airapetian:2004tw,
Alexakhin:2005iw,Ageev:2006da,Alekseev:2008dn,Airapetian:2009ti,
Avakian:2010ae,Alekseev:2010dm,Adams:1991rw,Adams:1991cs,Bravar:1996ki,Adams:2003fx}.

{}

In previous studies we discussed the covariant parton model, which is based
on the 3D picture of parton momenta with rotational symmetry in the nucleon 
rest frame \cite{Zavada:1996kp, Zavada:2001bq,
Zavada:2002uz, Efremov:2004tz, Zavada:2007ww, Efremov:2008mp, Efremov:2009ze,
Zavada:2009sk,Efremov:2009vb,Avakian:2010nz}. 
An important feature of this approach is the implication of relations among 
various PDFs, such as the Wandzura-Wilczek approximation between $g_{1}^q(x)$ 
and $g_{2}^q(x)$ which was proven in \cite{Zavada:2001bq} together with some 
other sum rules. Assuming $SU(6)$ symmetry 
(in addition to Lorentz invariance and rotational symmetry) 
relations between the polarized and unpolarized structure functions were found
\cite{Zavada:2002uz}, which agree very well with data. In \cite{Efremov:2004tz} 
transversity was studied in the framework of this model and a relation between
transversity $h_1^q(x)$ and helicity $g_1^q(x)$ was obtained. In a next step 
we generalized the model to the description of TMDs.
We derived a relation between the pretzelosity distribution $h_{1T}^{\perp a}$, 
transversity and helicity \cite{Efremov:2008mp}.
Finally, with the same model we studied all time-reversal even (T-even) TMDs 
and derived a set of relations among them \cite{Efremov:2009ze}. Moreover, 
it was also shown that the 3D picture of parton momenta inside the nucleon 
provides a basis for a consistent description of quark orbital angular momentum
\cite{Zavada:2007ww}, which is related to pretzelosity \cite{Avakian:2010nz}.
It should be remarked that some of the relations among different TMDs were found 
(sometimes before) also in other models 
\cite{Jakob:1997wg,Meissner:2007rx,Avakian:2007mv,Avakian:2008dz,
Pasquini:2008ax,She:2009jq,Avakian:2010br,Pasquini:2010pa}.

The comparison of the obtained relations and predictions with experimental data 
is very important and interesting from phenomenological point of view. It
allows us to judge to which extent the experimental observation can be
interpreted in terms of simplified, intuitive notions. The obtained picture
of the nucleon can be a useful supplement to the exact but more 
complicated theory of the nucleon structure based on QCD. For example,
the covariant parton model can be a useful tool for separating effects of 
QCD from effects of relativistic kinematics.

In this paper we further develop and broadly extend our studies 
\cite{Zavada:2009sk,Efremov:2009vb} of the relations between TMDs and PDFs. 
The formulation of the model in terms of the light-cone formalism 
\cite{Efremov:2009ze} allows us to compute the leading-twist TMDs 
by means of the light-front correlators $\phi (x,\mathbf{p}_{T})_{ij}$ 
\cite{Mulders:1995dh} as:%
\begin{eqnarray}
\frac{1}{2}\,\mathrm{tr}\left[ \gamma ^{+}\;\phi (x,\mathbf{p}_{T})\right]
&=&f_{1}^q(x,\mathbf{p}_{T})-\frac{\varepsilon ^{jk}p_{T}^{j}S_{T}^{k}}{M}%
\,f_{1T}^{\perp a}(x,\mathbf{p}_{T}),  \label{e1}\\
\frac{1}{2}\,\mathrm{tr}\left[ \gamma ^{+}\gamma _{5}\phi (x,\mathbf{p}_{T})\right] 
&=&S_{L}g_{1}^q(x,\mathbf{p}_{T})+\frac{\mathbf{p}_{T}\mathbf{S}_{T}}{M}%
g_{1T}^{\bot a}(x,\mathbf{p}_{T}),  \label{e2}\\
\hspace{-8mm}
\frac{1}{2}\,\mathrm{tr}\left[ i\sigma ^{j+}\gamma _{5}\phi (x,\mathbf{p}_{T})\right]  
&=&S_{T}^{j}\,h_{1}^q(x,\mathbf{p}_{T})
+S_{L}\,\frac{p_{T}^{j}}{M}\,h_{1L}^{\perp a}(x,\mathbf{p}_{T})  \label{ee2} 
+\frac{(p_{T}^{j}p_{T}^{k}-\frac{1}{2}\,\mathbf{p}_{T}^{2}
 \delta^{jk})S_{T}^{k}}{M^{2}}\,h_{1T}^{\perp a}(x,\mathbf{p}_{T})
+\frac{\varepsilon^{jk}p_{T}^{k}}{M}\,h_{1}^{\perp a}(x,\mathbf{p}_{T}).\;\;
\end{eqnarray}%
The main goal of this work is to derive relations between these TMDs and the 
usual PDFs $f_{1}(x)$ and $g_{1}(x)$. Similar tasks were recently addressed
also in other approaches \cite{D'Alesio:2009kv,Bourrely:2010ng}. 
The Section~\ref{App-1:unpolarized-TMDs} is devoted to the case of the
unpolarized TMD, and in Section~\ref{App-2:polarized-TMDs} we discuss
polarized TMDs. We use the obtained relations to calculate and discuss 
the numerical predictions for TMDs. The Section~\ref{sec4} is devoted 
to our concluding remarks.

\newpage
\section{The unpolarized TMD}

\label{App-1:unpolarized-TMDs}

The distribution  $f_{1}^{q}(x,\mathbf{p}_{T})$ is given  in the covariant
parton model approach  by the expression \cite{Efremov:2009ze} 
\begin{equation}
f_{1}^{q}(x,\mathbf{p}_{T})=xM\int \frac{\mathrm{d}p^{1}}{p^{0}}%
\;G(p^{0})\,\delta \left( \frac{p^{0}-p^{1}}{M}-x\right) =M\;G(\bar{p}%
^{0})\,.  \label{Eq:App-f1-1}
\end{equation}%
In the final step of (\ref{Eq:App-f1-1}) we performed the $p^{1}$%
-integration by rewriting the $\delta $-function as 
\begin{equation}
x\;\delta \left( \frac{p^{0}-p^{1}}{M}-x\right) = 
\bar{p}%
^{0}\,\delta (p^{1}-\bar{p}^{1})\;,\;\;\;\bar{p}^{0}=\frac{1}{2}\,xM\,\left(
1+\frac{\mathbf{p}_{T}^{2}+m^{2}}{x^{2}M^{2}}\right) ,\;\;\;\bar{p}^{1}=-%
\frac{1}{2}\,xM\,\left( 1-\frac{\mathbf{p}_{T}^{2}+m^{2}}{x^{2}M^{2}}\right).  
\label{Eq:identity-delta}
\end{equation}%
The remarkable feature of the present parton model approach is that one can
predict unambiguously the $x$- and $\mathbf{p}_{T}$-dependence of TMDs from 
the $x$-dependence of the corresponding (\textquotedblleft
integrated\textquotedblright ) parton distribution functions. The deeper
reason for that is the equal (3D-symmetric in the nucleon rest frame)
description of longitudinal (i.e. $p^{1}$-dependence) and transverse ($%
\mathbf{p}_{T}$-dependence) parton momenta. Let us remark, that the
invariant parameter $x$ (Bjorken $x$) is tightly connected to both
longitudinal and transverse parton through the $\delta $-function.
In the nucleon rest frame transverse momenta play for $x$ an
important role according to (\ref{Eq:identity-delta}). 

The 3D momentum distribution $G^{q}(p^{0})$ was expressed in terms of
$f_1^q(x)$ in previous works \cite{Zavada:2001bq,Zavada:2007ww}. 
In order to make this work self-contained we present here an 
independent derivation. We start from the model expression for 
$f_{1}^{q}(x)$ which follows from the first equality in Eq.~(\ref{Eq:App-f1-1}). 
For the remainder of this section we
set the parton mass $m\rightarrow 0$. Besides being a reasonable
approximation, this step greatly simplifies the calculation though finite $m$%
-effect can be included \cite{Zavada:2002uz}. Notice that if $m$ is
neglected then $p^{0}=\sqrt{p_{1}^{2}+\mathbf{p}_{T}^{2}}\equiv p$. Now, instead
of integrating over $p^{1}$ as we did in Eq.~(\ref{Eq:App-f1-1}), it is
convenient to use spherical coordinates, and define the angles such that 
$p^{1}=p\,\cos \theta $, i.e.\ 
\begin{eqnarray}
f_{1}^{q}(x) &=&xM\int \frac{\mathrm{d}^{3}p}{p}\;G^{q}(p)\,\delta \left( 
\frac{p-p\,\cos \theta }{M}-x\right)   \nonumber \\
&=&xM\int_{0}^{2\pi }\mathrm{d}\phi \int_{0}^{\infty }\mathrm{d}%
p\;p\,G^{q}(p)\,\int_{-1}^{1}\mathrm{d}\cos \theta \;\delta \left( \frac{%
p-p\,\cos \theta }{M}-x\right)   \nonumber \\
&=&2\pi \,xM^{2}\int_{0}^{\infty }\mathrm{d}p\;G^{q}(p)\;\Theta \left( p-%
\frac{1}{2}\,xM\right) .  \label{Eq:inverting-f1}
\end{eqnarray}%
The $\Theta $-function emerges because the integral over the $\delta$-function
obviously yields a non-zero result only if $|\cos \theta |<1$, and implies a lower 
limit for $p$-integral. Notice that there is also an upper limit, namely 
$p<\frac{1}{2}M$,  related to the fact that $x<1$ \cite{Zavada:2001bq}.
This upper limit is natural in the covariant parton model in the nucleon
rest frame because an on-shell parton can carry at most the momentum $p_{%
\mathrm{max}}=\frac{1}{2}M$ which must be compensated by all other partons
going in the opposite direction, such that the center of mass of the nucleon
remains at rest. Since this is an unlikely constellation the momentum
distribution $G^{q}(p)$ vanishes as $p\rightarrow p_{\mathrm{max}}$
similarly as $f_{1}^{q}(x)$ drops to zero with $x\rightarrow 1$. Thus we
obtain 
\begin{equation}
\frac{f_{1}^{q}(x)}{x}=2\pi M^{2}\int\limits_{\frac{1}{2}\,xM}^{\frac{1}{2}%
\,M}\mathrm{d}p\;G^{q}(p)  \label{Eq:f1-Gp-integ}
\end{equation}%
and we reproduce the identity  \cite{Zavada:2001bq,Zavada:2007ww} 
\begin{equation}
\frac{\mathrm{d}\;}{\mathrm{d}x}\left[ \frac{f_{1}^{q}(x)}{x}\right] =-\pi
M^{3}\;G^{q}\left( \frac{xM}{2}\right) .  \label{Eq:f1-diff-Gp}
\end{equation}%
This result inserted in (\ref{Eq:App-f1-1}) enables us to predict uniquely 
$f_{1}^{q}(x,\mathbf{p}_{T})$ from $f_{1}^{q}(x)$ as follows 
\begin{equation}
f_{1}^{q}(x,\mathbf{p}_{T})=-\;\frac{1}{\pi M^{2}}\;\frac{\mathrm{d}\;}{%
\mathrm{d}y}\left[ \frac{f_{1}^{q}(y)}{y}\right] _{y=\xi (x,\mathbf{p}%
_{T}^{2})} \label{Eq:f1-pT-predicted}
\end{equation}
with the dependence on $x$, ${\bf p}_T$ given through the variable 

\begin{equation}
\xi (x,\mathbf{p}_{T}^{2})=\lim\limits_{m%
\rightarrow 0}\frac{2\bar{p}^{0}}{M}=\,x\,\left( 1+\frac{\mathbf{p}_{T}^{2}}{%
x^{2}M^{2}}\right) \;.  \label{Eq:def-xi}
\end{equation}
The variable $\xi (x,\mathbf{p}_{T}^{2})$, first suggested in
\cite{Zavada:1996kp}, relates the (very different!) dependencies on 
longitudinal and transverse momenta so that the factorization of these
dependencies can be only approximate. It is in some sense similar to the
Nachtmann variable in DIS which controls some (kinematical) part of higher
twist contributions. The reason for the apperance of such a variable is 
deeply related to the general properties of our model like Lorentz invariance
\cite{Zavada:1996kp,D'Alesio:2009kv} and the on-shellness of partons.

\begin{figure}[tbp]
\includegraphics[width=12cm]{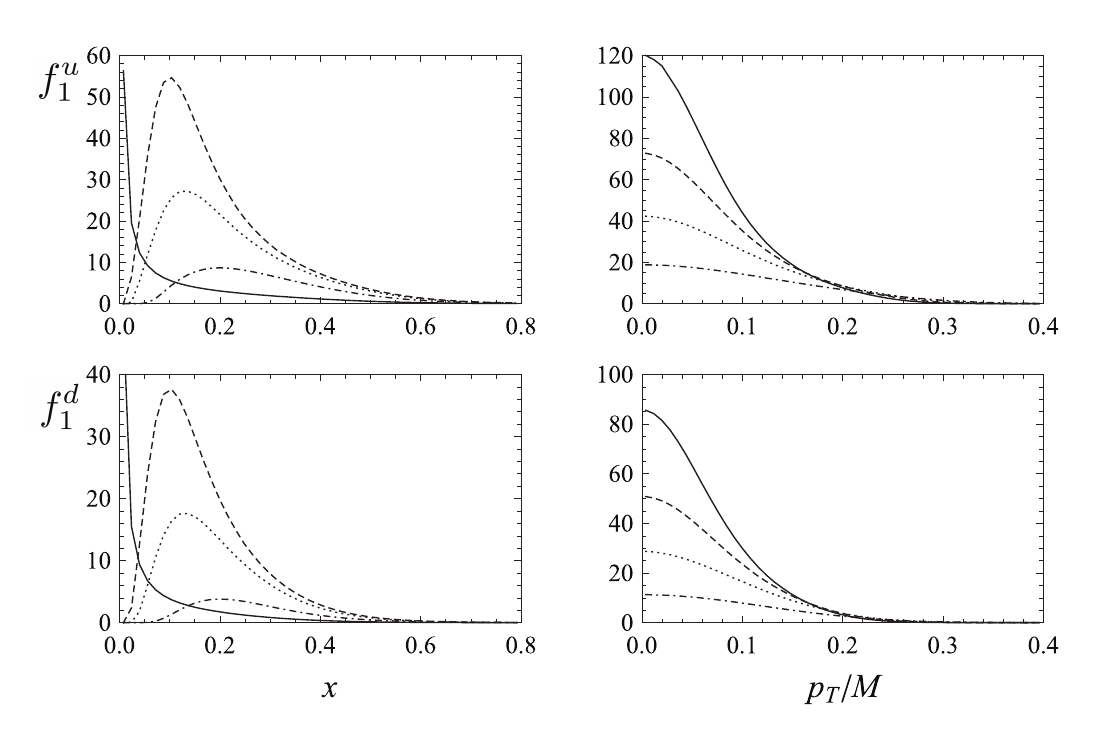}

\vspace{-3mm}

\caption{The TMDs $f_1^q(x,{\bf p}_T)$ for $u$- (\textit{upper panel}) and $d$-quarks 
(\textit{lower panel}).
\textbf{Left panel}:  $f_1^q(x,{\bf p}_T)$ as function of $x$ for $p_{T}/M=0.10$ (dashed),
0.13 (dotted), 0.20 (dashed-dotted line). The solid line corresponds to the input
distribution $f_{1}^{q}(x)$. 
\textbf{Right panel}: $f_1^q(x,{\bf p}_T)$ as function of $p_{T}/M$ for 
$x=0.15$ (solid), 0.18 (dashed), 0.22 (dotted), 0.30 (dashed-dot line).}
\label{ff1}
\end{figure}

\begin{figure}[tbp]
\includegraphics[width=12cm]{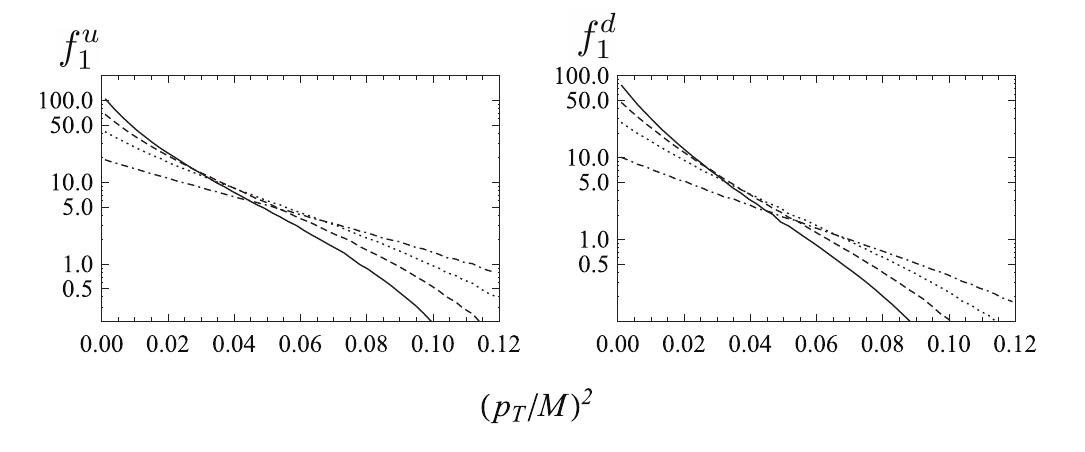}

\vspace{-3mm}

\caption{$f_1^q(x,{\bf p}_T)$ as function of $(p_{T}/M)^2$ for 
$x=0.15$ (solid), 0.18 (dashed), 0.22 (dotted), 0.30 (dash-dotted line).}
\label{ff2}
\end{figure}

Using as input for $f_{1}^{q}(x)$ the LO parameterization of \cite{Martin:2004dh} 
at the scale $4\,{\rm GeV}^{2}$, we obtain for $u$ and $d$-quarks the results shown
in Fig.~\ref{ff1}.
The lower part of this figure is shown again, in a different scale in Fig.~\ref{ff2}. 

\noindent
We make the following observations:

\textit{i)} For fixed $x$, the $p_{T}$-distributions are similar to the
Gauss Ansatz $f_{1}^{q}(x,p_{T})\propto \exp \left( -p_{T}^{2}/\left\langle
p_{T}^{2}\right\rangle \right)$. This is an interesting result, since the
Gaussian shape is supported by phenomenology \cite{Schweitzer:2010tt}.

\textit{ii)} The width $\left\langle p_{T}^{2}\right\rangle $ depends on $x$.
This result reflects the fact, that in our approach the parameters $x$ and 
$p_{T}$ are not independent due to rotational symmetry.

\textit{iii)} The Figs.~\ref{ff1},~\ref{ff2} suggest that the typical values 
for transverse momenta, $\left\langle p_{T}^{2}\right\rangle \approx 0.01GeV^{2}$ 
or $\left\langle p_{T}\right\rangle \approx 0.1GeV$. These values correspond 
to the estimates based on the different analyses of the structure function 
$F_{2}(x,Q^{2})$ \cite{Zavada:2009sk}. On the other hand, much larger values 
$\langle p_{T}^{2}\rangle \sim 0.4GeV^{2}$ are inferred from SIDIS data 
referring to comparable scales \cite{Schweitzer:2010tt,
Anselmino:2005nn,Collins:2005ie}. Note also that in the statistical model
of TMDs \cite{Bourrely:2005tp,Bourrely:2010ng} the parameter $\left\langle
p_{T}\right\rangle $ may be interpreted as an effective temperature of the 
partonic "ensemble" \cite{Cleymans:2010aa}. It is instructive to compare this
number to the lattice calculations \cite{Karsch:2001vs} of the QCD phase
transition temperature $T\approx 175$ MeV.

\section{The polarized TMDs}

\label{App-2:polarized-TMDs}

All polarized leading-twist T-even TMDs 
are described in terms of the \textsl{same} polarized covariant 3D
distribution $H(p^{0})$. This follows from the compliance of the approach
with relations following from QCD equations of motion \cite{Efremov:2009ze}.
As a consequence all polarized TMDs can be expressed in terms a
single \textquotedblleft generating function\textquotedblright\ $K^{q}(x,%
\mathbf{p}_{T})$ as follows 
\begin{equation}
\renewcommand{\arraystretch}{2.2}
\begin{array}{rcrcl}
g_{1}^{q}(x,\mathbf{p}_{T}) & = & \displaystyle \frac{1}{2x}\left( \left( x+\frac{m}{M}%
\right) ^{2}-\frac{\mathbf{p}_{T}^{2}}{M^{2}}\right)  & \times  & K^{q}(x,%
\mathbf{p}_{T})\;, \\ 
h_{1}^{q}(x,\mathbf{p}_{T}) & = & \displaystyle \frac{1}{2x}\left( x+\frac{m}{M}\right)
^{2} & \times  & K^{q}(x,\mathbf{p}_{T})\;, \\  
g_{1T}^{\perp q}(x,\mathbf{p}_{T}) & = & \displaystyle \frac{1}{x}\left( x+\frac{m}{M}%
\right) \; & \times  & K^{q}(x,\mathbf{p}_{T})\;, \\ 
h_{1L}^{\perp q}(x,\mathbf{p}_{T}) & = & \displaystyle -\,\frac{1}{x}\left( x+\frac{m}{M}%
\right) \; & \times  & K^{q}(x,\mathbf{p}_{T})\;, \\ 
h_{1T}^{\perp q}(x,\mathbf{p}_{T}) & = & \displaystyle -\frac{1}{x}\, & \times  & K^{q}(x,%
\mathbf{p}_{T})\;.%
\end{array}
\label{Eq:all-TMDs}
\end{equation}%
with the \textquotedblleft generating function\textquotedblright\ $K^{q}(x,%
\mathbf{p}_{T})$ defined (in the compact notation introduced in Eq.~(32) of 
\cite{Efremov:2009ze}) by 
\begin{equation}
K^{q}(x,\mathbf{p}_{T})=M^{2}x\int \mathrm{d}\{p^{1}\}\;\;,\;\;\;\;\mathrm{d}%
\{p^{1}\}\equiv \frac{\mathrm{d}p^{1}}{p^{0}}\;\frac{H^{q}(p^{0})}{p^{0}+m}%
\;\delta \left( \frac{p^{0}-p^{1}}{M}-x\right) \,.
\label{Eq:generating-function}
\end{equation}
We remark that in order to rewrite $g_{1}^{q}(x,%
\mathbf{p}_{T})=\int \mathrm{d}\{p^{1}\}\,[(p^{0}+m)xM-\mathbf{p}_{T}^{2}]$ 
\cite{Efremov:2009ze} as shown in the first equation of (\ref{Eq:all-TMDs}), we
used the identity $p^{0}=(\mathbf{p}_{T}^{2}+x^{2}M^{2}+m^{2})/(2xM)$ valid
under the $p^{1}$-integral, which holds because in the model the partons are
on-shell, i.e.\ $p_{0}^{2}-p_{1}^{2}-\mathbf{p}_{T}^{2}=m^{2}$, and $%
p^{0}-p^{1}=xM$ due to the delta-function. The expressions for the other
TMDs in (\ref{Eq:all-TMDs}) can be read off directly from Eqs.~(35-38) in 
\cite{Efremov:2009ze}. Using the identity (\ref{Eq:identity-delta}) we
perform the $p^{1}$-integration in (\ref{Eq:generating-function}) and obtain
for the generating function, which depends on $x$, $\mathbf{p}_{T}$ only via
$\bar{p}^{0}$ defined in Eq. (\ref{Eq:identity-delta}):  
\begin{equation}
K^{q}(x,\mathbf{p}_{T})=M^{2}\frac{H^{q}(\bar{p}^{0})}{\bar{p}^{0}+m},\qquad 
\bar{p}^{0}=\frac{1}{2}\,xM\,\left( 1+\frac{\mathbf{p}_{T}^{2}+m^{2}}{%
x^{2}M^{2}}\right) .  \label{Eq:generating-function-2}
\end{equation}

From (\ref{Eq:all-TMDs}--\ref{Eq:generating-function-2}) it is clear that we
can predict all polarized TMDs if we know $H^{q}(p^{0})$. The polarized 3D
momentum distribution $H^{q}(p^{0})$ could be determined in principle from
any polarized TMD, but the helicity parton distribution function $%
g_{1}^{q}(x)$ plays a special role, because its $x$-dependence is known. The
connection of $g_{1}^{q}(x)$ and $H^{q}(p^{0})$ was derived previously in 
\cite{Zavada:2001bq,Zavada:2007ww}. In order to make this work
self-contained we present here an independent derivation.

We start from the expression for $g_{1}^{q}(x)$ which follows from (\ref%
{Eq:all-TMDs}) and proceed as in Eq.~(\ref{Eq:inverting-f1}), i.e.\ we
neglect $m$ and use spherical coordinates such that $p^{1}=p\,\cos \theta $
and $\mathbf{p}_{T}^{2}=p^{2}\,\sin ^{2}\theta $, i.e.\ 
\begin{eqnarray}
g_{1}^{q}(x) &=&\int \frac{\mathrm{d}^{3}p}{2\,p^{2}}\;H^{q}(p)%
\,(x^{2}M^{2}-p^{2}\sin ^{2}\theta )\;\delta \left( \frac{p-p\,\cos \theta }{%
M}-x\right)   \label{Eq:g1-starting} \\
&=&\int_{0}^{2\pi }\mathrm{d}\phi \int \frac{\mathrm{d}p}{2}%
\;H^{q}(p)\,\int_{-1}^{1}\mathrm{d}\cos \theta \;(x^{2}M^{2}-p^{2}\sin
^{2}\theta )\;\delta \left( \frac{p-p\,\cos \theta }{M}-x\right)   \nonumber
\\
&=&2\pi M^{2}\int \mathrm{d}p\;H^{q}(p)\;\frac{x}{p}(xM-p)\;\Theta \left( p-%
\frac{1}{2}\,xM\right)   \nonumber
\end{eqnarray}%
where the $\Theta $-function emerges in the same way it did in Eq.~(\ref{Eq:inverting-f1}). 
Consequently, we obtain%
\begin{equation}
g_{1}^{q}(x)=2\pi \,M^{2}x\;\left( xM\int\limits_{\frac{1}{2}xM}^{\frac{1}{2}%
M}\frac{\mathrm{d}p}{p}\,H^{q}(p)-\int\limits_{\frac{1}{2}xM}^{\frac{1}{2}M}%
\mathrm{d}p\;H^{q}(p)\right) \;.  \label{Eq:identities-with-g1a}
\end{equation}%
Differentiating this equation and integrating it by parts gives the
following results 
\begin{eqnarray}
x\;\frac{\mathrm{d}g_{1}^{q}(x)}{\mathrm{d}x} &=&2\pi \,M^{2}x\left(
2\,xM\int\limits_{\frac{1}{2}xM}^{\frac{1}{2}M}\frac{\mathrm{d}p}{p}%
\,H^{q}(p)-\int\limits_{\frac{1}{2}xM}^{\frac{1}{2}M}\mathrm{d}p\;H^{q}(p)-%
\frac{xM}{2}H^{q}\!\left( \frac{xM}{2}\right) \right) \;,  \nonumber \\
\int_{x}^{1}\frac{\mathrm{d}y}{y}\;g_{1}^{q}(y) &=&2\pi \,M^{2}x\left( \frac{%
-xM}{2}\int\limits_{\frac{1}{2}xM}^{\frac{1}{2}M}\frac{\mathrm{d}p}{p}%
\,H^{q}(p)+\int\limits_{\frac{1}{2}xM}^{\frac{1}{2}M}\mathrm{d}%
p\;H^{q}(p)\right) \;.  \label{Eq:identities-with-g1}
\end{eqnarray}%
Now we see that if we take the linear combination $2\int_{x}^{1}\mathrm{d}%
y\,g_{1}^{q}(y)/y+3\,g_{1}^{q}(x)-x\,g_{1}^{q\,\prime }(x)$ the integral
terms from the last three equations cancel out, and we obtain for 
$p^{2}H^{q}(p)$ at $p=\frac{M}{2}x$ the expression
\begin{equation}
\pi x^{2}M^{3}H^{q}\!\left( \frac{M}{2}x\right) =2\int_{x}^{1}\frac{\mathrm{d%
}y}{y}\;g_{1}^{q}(y)+3\,g_{1}^{q}(x)-x\;\frac{\mathrm{d}g_{1}^{q}(x)}{%
\mathrm{d}x},  \label{Eq:relation-g1-Hp}
\end{equation}%
which confirms previous works \cite{Zavada:2001bq,Zavada:2007ww}. 
For the generating function (\ref{Eq:generating-function-2}) we obtain,
in the limit $m\rightarrow 0$, the result
\begin{equation}
K^{q}(x,\mathbf{p}_{T})=\frac{H^{q}(\frac{M}{2}\xi )}{\,\frac{M}{2}\xi }=%
\frac{2}{\pi \xi ^{3}M^{4}}\left( 2\int_{\xi }^{1}\frac{\mathrm{d}y}{y}%
\;g_{1}^{q}(y)+3\,g_{1}^{q}(\xi )-x\;\frac{\mathrm{d}g_{1}^{q}(\xi )}{%
\mathrm{d}\xi }\right) ,\qquad \xi =\,x\,\left( 1+\frac{\mathbf{p}_{T}^{2}}{%
x^{2}M^{2}}\right) .  \label{e3}
\end{equation}%
and from (\ref{Eq:all-TMDs}) we obtain 
(in agreement with the result reported in the proceeding \cite{Efremov:2009vb}
which was derived independently)
\begin{equation}
g_{1}^{q}(x,\mathbf{p}_{T})=\frac{2x-\xi }{\pi \xi ^{3}M^{3}}\left(
2\int_{\xi }^{1}\frac{\mathrm{d}y}{y}\;g_{1}^{q}(y)+3\,g_{1}^{q}(\xi )-\xi \;%
\frac{\mathrm{d}g_{1}^{q}(\xi )}{\mathrm{d}\xi }\right) .  \label{e4}
\end{equation}%
In (\ref{e3},~\ref{e4}) and also below in (\ref{m23}) we use the 
variable $\xi=\xi(x,{\bf p}_T)$ as defined in (\ref{Eq:def-xi}).

\begin{figure}[tbp]
\includegraphics[width=12cm]{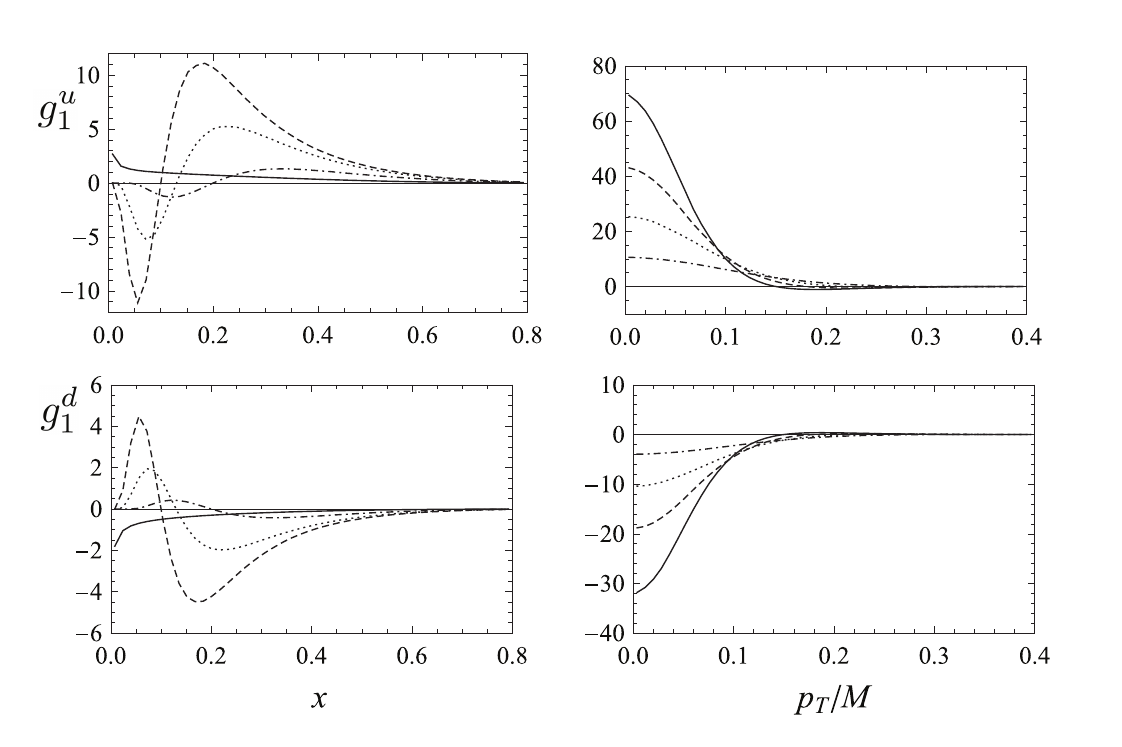}

\vspace{-3mm}

\caption{The TMD $g_1^q(x,{\bf p_T})$ for $u$- (\textit{upper panel}) and $d$-quarks 
(\textit{lower panel}). \textbf{Left panel}: $g_1^q(x,{\bf p_T})$ as function of $x$ 
for $p_{T}/M=0.10$ (dashed), 0.13 (dotted), 0.20 (dash-dotted line). The solid 
line corresponds to the input distribution $g_{1}^{q}(x)$. \textbf{Right panel}: 
$g_1^q(x,{\bf p_T})$ as function of $p_{T}/M$  for 
$x=0.15$ (solid), 0.18 (dashed), 0.22 (dotted), 0.30 (dash-dotted line).}
\label{ff3}
\end{figure}

Eq.~(\ref{e4}) yields for $g_{1}^{q}(x,{\bf p}_{T})$,  
with the LO parameterization of \cite{lss} for $g_1^q(x)$ at $4\,{\rm GeV}^2$, 
the results shown in Fig.~\ref{ff3}. 
\begin{figure}[tbp]
\includegraphics[width=12cm]{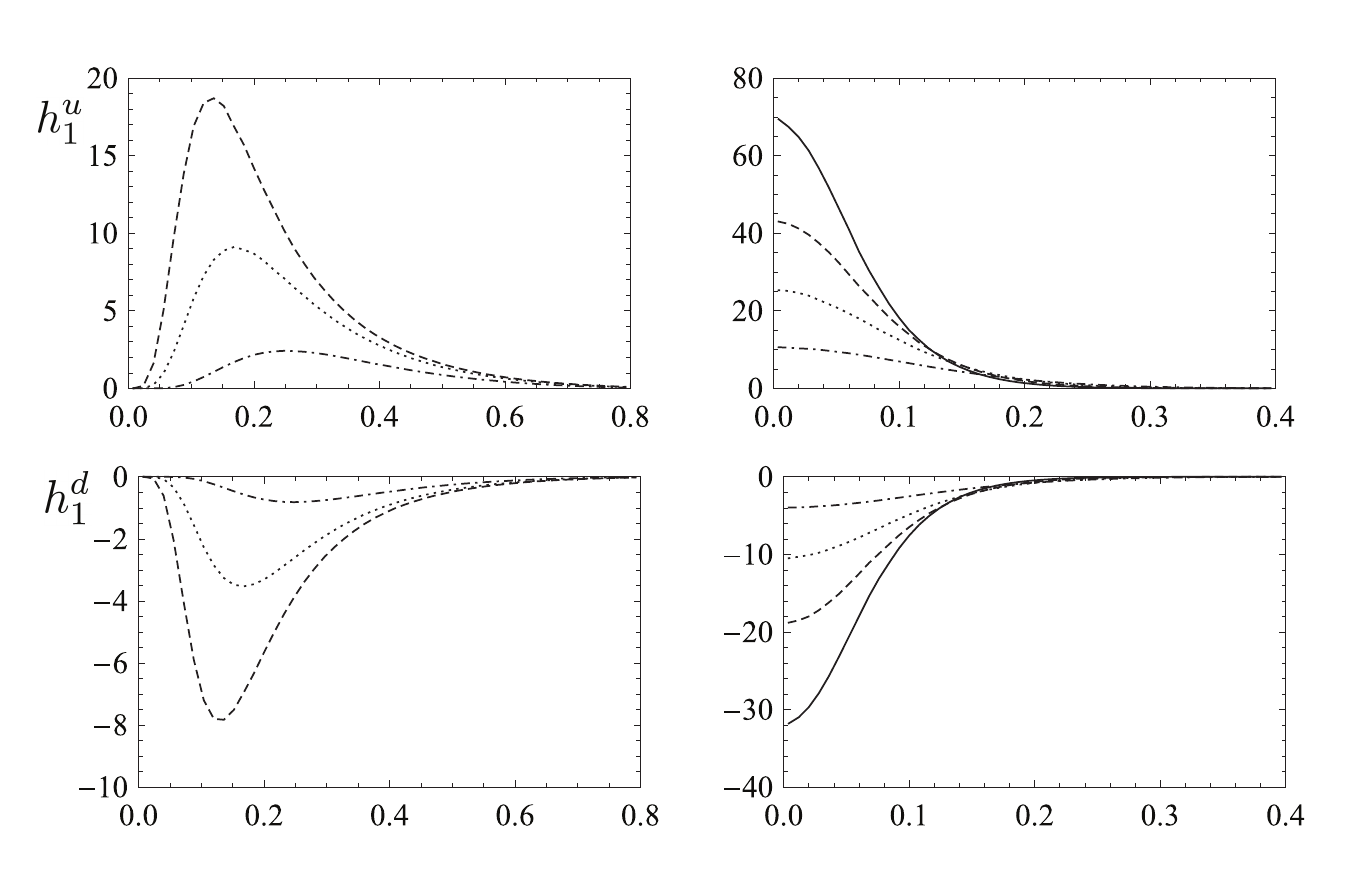} 
\includegraphics[width=12cm]{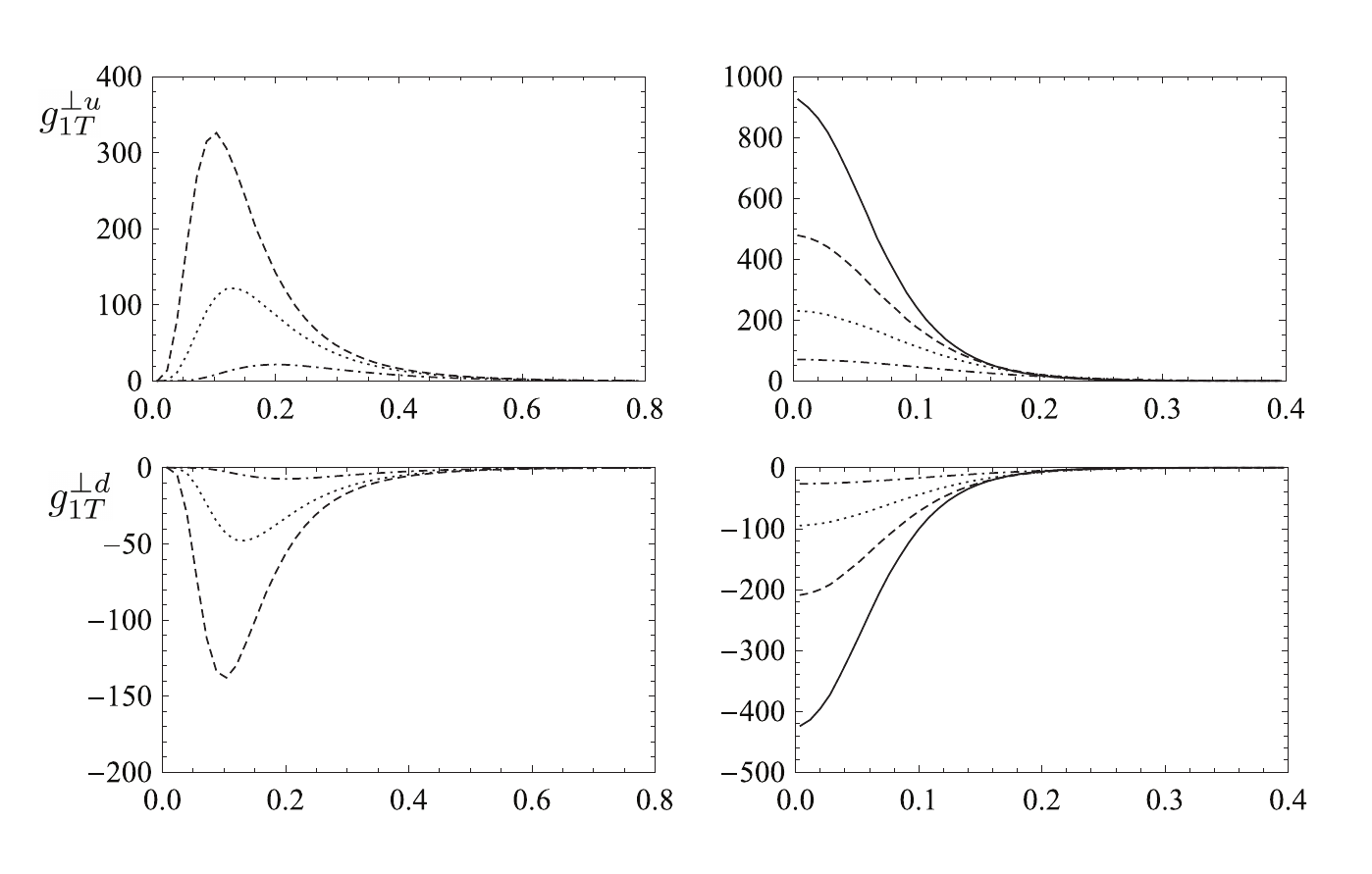} %
\includegraphics[width=12cm]{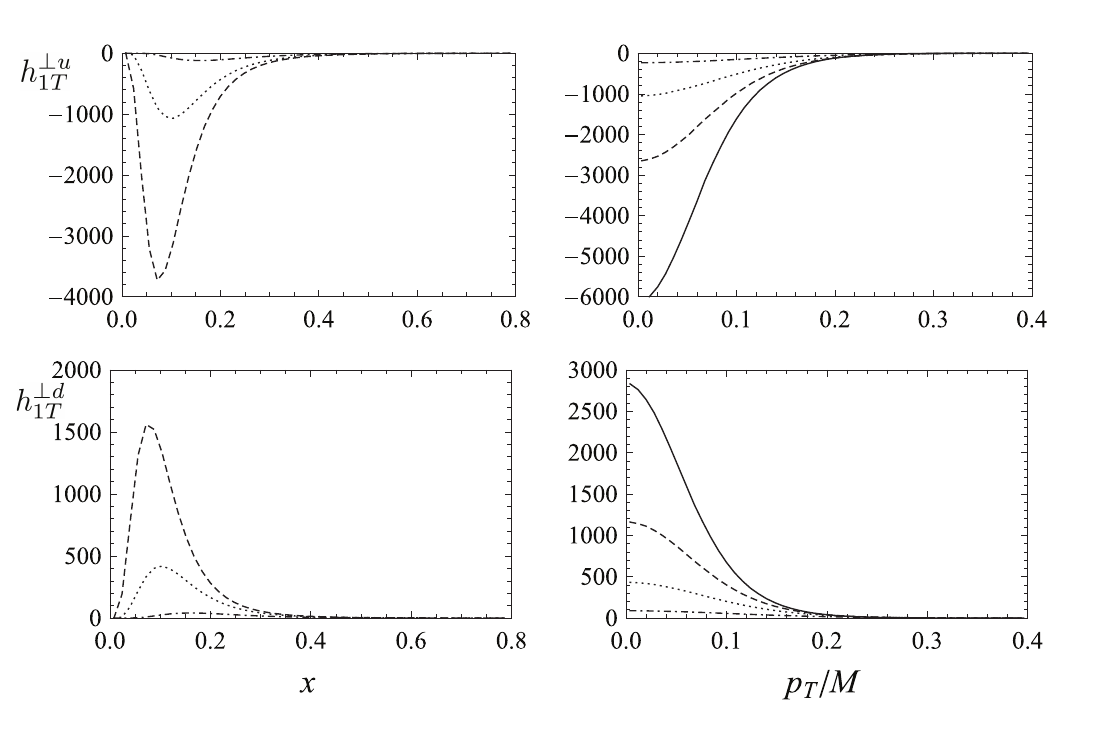}
\caption{\label{ff4}The TMDs $h_{1}^{q}(x,{\bf p_T})$, $g_{1T}^{\bot q}(x,{\bf p_T})$, 
$h_{1T}^{\perp q}(x,{\bf p_T})$ for $u$- and $d$-quarks.
\textbf{\ Left panel}: The TMDs as functions of $x$ for 
$p_{T}/M=0.10$ (dashed), 0.13 (dotted), 0.20(dash-dotted lines). 
\textbf{Right panel}: The TMDs as functions of $p_{T}/M$ for 
$x=0.15$ (solid), 0.18 (dashed), 0.22 (dotted), 0.30 (dash-dotted lines).}
\end{figure}
The remarkable observation is that $g_1^q(x,{\bf p}_T)$ changes sign at the 
point $p_{T}=Mx$, which is due to the prefactor 
\begin{equation}
2x-\xi =x\left( 1-\left( \frac{p_{T}}{Mx}\right) ^{2}\right) =-2\bar{p}^{1}/M
\label{m23}
\end{equation}%
in (\ref{e4}). 
The expression in (\ref{m23}) is proportional 
to the quark longitudinal
momentum $\bar{p}^{1}$ in the proton rest frame, which is determined 
by $x$ and $p_{T}$, see Eq. (\ref{Eq:identity-delta}). This means, that the 
sign of $g_{1}^{q}(x,p_{T})$ is controlled by sign of $\bar{p}^{1}$. 
To observe these dramatic sign changes one may look for multi-hadron
jet-like final states in SIDIS. Performing the cutoff for transverse
momenta from below and from above, respectively, should effect the sign of
asymmetry.

There is some similarity to $g_{2}^{q}(x)$ which also changes sign,
and is given in the model by the expression \cite{Zavada:2007ww}
\begin{equation}
g_{2}^{q}(x)=\frac{1}{2}\int H^{q}(p^{0})\left( p^{1}-\frac{\left(
p^{1}\right) ^{2}-p_{T}^{2}/2}{p^{0}+m}\right) \:\delta \left( \frac{%
p^{0}-p^{1}}{M}-x\right) \frac{d^{3}p}{p^{0}}.  \label{sp11}
\end{equation}%
The $\delta -$function implies that, for our choice of the light-cone direction,
large $x$ are correlated with large and negative $p^{1}$, 
while low $x$ are correlated with large and positive $p^{1}$. 
Thus, $g_{2}(x)$\ changes sign, because the integrand in
(\ref{sp11}) changes sign between the extreme values of $p^{1}$.
Let us remark, that the calculation of $g_{2}(x)$ based on the relation 
(\ref{sp11}) well agrees \cite{Zavada:2002uz} with the experimental data.

The other TMDs (\ref{Eq:all-TMDs}) can be calculated similarly 
and differ, in the limit $m\to 0$, by simple $x$-dependent prefactors
\begin{eqnarray}
h_{1}^{q}(x,\mathbf{p}_{T}) &=&\frac{x}{2}K^{q}(x,\mathbf{p}_{T})\;,
\label{e5} \\
g_{1T}^{\perp q}(x,\mathbf{p}_{T}) &=&K^{q}(x,\mathbf{p}_{T})\;,  \nonumber
\\
h_{1L}^{\perp q}(x,\mathbf{p}_{T}) &=&-K^{q}(x,\mathbf{p}_{T})\;,  \nonumber
\\
h_{1T}^{\perp q}(x,\mathbf{p}_{T}) &=&-\frac{1}{x}K^{q}(x,\mathbf{p}_{T})\;.
\nonumber
\end{eqnarray}%
The resulting plots are shown in Fig.~\ref{ff4}. We do not plot
$h_{1L}^{\perp q}$ since this TMD is equal to $-g_{1T}^{\perp q}$ in our
approach \cite{Efremov:2009ze}. Let us remark, that $g_{1}^{q}(x,\mathbf{p}_{T})$
is the only TMD which can change sign. The other TMDs have all definite signs,
which follows from (\ref{Eq:all-TMDs},~\ref{e5}). Note also that pretzelosity 
$h_{1T}^{\perp q}(x,\mathbf{p}_{T})$, due to the prefactor $1/x$, has the largest 
absolute value among all TMDs.

\section{Concluding remarks}
\label{sec4}

We have studied relations between the TMDs $f_{1}^{q}(x,\mathbf{p}%
_{T}),g_{1}^{q}(x,\mathbf{p}_{T}),h_{1}^{q}(x,\mathbf{p}_{T}),g_{1T}^{\bot
q}(x,\mathbf{p}_{T}),h_{1L}^{\perp q}(x,\mathbf{p}_{T}),h_{1T}^{\perp q}(x,%
\mathbf{p}_{T})$ and the PDFs $f_{1}^{q}(x),g_{1}^{q}(x)$. 
These relations follow in the covariant parton model as 
a consequence of Lorentz invariance and rotational symmetry of
parton momenta in the nucleon rest frame. These relations used with 
parameterizations of PDFs give numerical predictions for all T-even
(in this work: leading twist) TMDs. In fact we have demonstrated, that in 
the framework of the model there are fundamental distributions covariant
momentum distributions $G^{q}(p)$ and $H^{q}(p)$ from which all the 
considered unpolarized and polarized PDFs and TMDs can be uniquely obtained. 
And vice versa, from any PDF or TMD one can unambiguously calculate
the corresponding $G^{q}$ or $H^{q}$. 

Some of our results are compatible with the results of the recent paper 
\cite{D'Alesio:2009kv}. In spite of some differences, both approaches have
an important common basis consisting in the Lorentz invariance. For a more
detailed comparison of the two approaches we refer to \cite{Zavada:2009sk}. 
Our predictions are consistent also with the results obtained in the recent
study \cite{Bourrely:2010ng}, some quantitative differences between these
two approaches are discussed in the cited paper.

To conclude, let us remark that an experimental check of the predicted TMDs
requires care. In fact, TMDs are not directly measurable 
quantities unlike structure functions. What one can measure for instance in
semi-inclusive DIS is a convolution with a quark fragmentation function. 
This naturally ``dilutes'' the effects of TMDs, and makes it difficult 
to observe for instance the prominent sign change in the helicity 
distribution, see Fig. \ref{ff3}. A dedicated study of the phenomenological
implications of our results is in progress. 

\vspace{3mm}

\noindent \textbf{Acknowledgements.} A.~E. and O.~T. are supported by the
Grants RFBR 09-02-01149 and 09-02-00732, and
(also P.Z.) Votruba-Blokhitsev Programs of JINR. P.~Z. is supported by the
project AV0Z10100502 of the Academy of Sciences of the Czech Republic. The
work was supported in part by DOE contract DE-AC05-06OR23177. We would like
to thank also Jacques Soffer and Claude Bourrely for helpful comments on
an earlier version of the manuscript.

\appendix

\end{document}